\documentclass{article}
\usepackage{spconf,amsmath,graphicx,amssymb,amsfonts,bm,algorithm, algorithmic,booktabs}
\usepackage[dvipsnames]{xcolor}
\usepackage[bibstyle=ieee,citestyle=numeric]{biblatex}
\usepackage{siunitx}
\usepackage{subcaption}
\usepackage{dsfont}
\usepackage[normalem]{ulem}
\usepackage[stretch=40,shrink=40]{microtype}
\DeclareSIUnit\byte{B}
\DeclareSIUnit\kilobyte{kB}
\DeclareSIUnit\megabyte{MB}
\DeclareSIUnit\gigabyte{GB}
\DeclareSIUnit\terabyte{TB}
\usepackage{svg}
\DeclareSIUnit{\million}{million}

\newcommand{\real}{\mathbb{R}}          
\newcommand{\trans}{^{\text{T}}}		

\title{Compressed BC-LISTA via Low-Rank Convolutional Decomposition}
\name{Han Wang\textsuperscript{*,\ddag}, Yhonatan Kvich\textsuperscript{\dag}, Eduardo P\'{e}rez\textsuperscript{*,\ddag},  Florian R\"{o}mer\textsuperscript{*,\ddag}, Yonina C. Eldar\textsuperscript{\dag}
\thanks{This research was supported by the Thuringian Ministry of Economic Affairs, Science and Digital Society (TMWWDG), and funded by the European Research Council (ERC) under the European Union’s Horizon 2020 research and innovation program (grant No. 101000967-CoDeS and 101119062-SWIMS), as well as by the Israel Science Foundation (grant No. 536/22).}
}
\address{
\textsuperscript{*}Ilmenau University of Technology, Ilmenau, Germany \\
\textsuperscript{\dag}Weizmann Institute of Science, Rehovot, Israel \\
\textsuperscript{\ddag}Fraunhofer Institute for Nondestructive Testing, Saarbrücken, Germany}

\begin{document}

\maketitle
\ninept

\begin{abstract}
We study Sparse Signal Recovery (SSR) methods for multichannel imaging with compressed {forward and backward} operators that preserve reconstruction accuracy. We propose a Compressed Block-Convolutional (C-BC) measurement model based on a low-rank Convolutional Neural Network (CNN) decomposition that is analytically initialized from a low-rank factorization of physics-derived forward/backward operators in time delay-based measurements. We use Orthogonal Matching Pursuit (OMP) to select a compact set of basis filters from the analytic model and compute linear mixing coefficients to approximate the full model. We consider the Learned Iterative Shrinkage-Thresholding Algorithm (LISTA) network as a representative example for which the C-BC-LISTA extension is presented. In simulated multichannel ultrasound imaging across multiple Signal-to-Noise Ratios (SNRs), C-BC-LISTA requires substantially fewer parameters and smaller model size than other state-of-the-art (SOTA) methods while improving reconstruction accuracy. In ablations over OMP, Singular Value Decomposition (SVD)-based, and random initializations, OMP-initialized structured compression performs best, yielding the most efficient training and the best performance.

\end{abstract}


\vspace{-1.0em}
\section{Introduction and Related Work}
\vspace{-0.5em}
Sparse Signal Recovery (SSR) and Compressed Sensing (CS) \cite{donoho2006compressed, eldar2012compressed, eldar2015sampling} address the problem of linear inversion under sparsity constraints. Let a linear measurement process $\mathbf{y=Ax}$ consist of a sparse signal of interest $\mathbf{x}\in \mathbb{R}^{N_\mathrm{s}}$, data $\mathbf{y}\in \mathbb{R}^{N_\mathrm{d}}$, and a forward model matrix $\mathbf{A}\in \mathbb{R}^{N_\mathrm{d}\times N_\mathrm{s}}$. 
Typically, in CS, we have $N_\mathrm{d}\ll N_\mathrm{s}$, which makes the problem ill-posed; even in less common cases such as fully sampled but rank-deficient acquisitions ($\operatorname{rank}(\mathbf{A})<N_\mathrm{s}$), the inverse problem remains ill-posed \cite{hansen1998rank}.
To recover $\mathbf{x}$, common algorithms ranging from classical sparsity-based methods to modern generative priors \cite{bora2017compressed} rely on iterative application of a forward and a backward model. 
As a representative example, a single ISTA step is formulated as:
\begin{equation} \label{eq_ista}
    \mathbf{x}^{(k+1)}=\mathcal{S}_{\frac{\lambda}{L}}\left(\left(\mathbf{I}-\frac{1}{L}\mathbf{A}\trans\mathbf{A}\right)\mathbf{x}^{(k)}+\frac{1}{L}\mathbf{A}\trans\mathbf{y}\right),
\end{equation}
where $\mathcal{S}_{\frac{\lambda}{L}}(\cdot)$ is the shrinkage-thresholding operator with the parameter $\frac{\lambda}{L}$, $L$ is the Lipschitz constant of $\mathbf{A}$, $\lambda$ is the regularization weight on the $\ell_1$ term in the LASSO problem, and $(\cdot)\trans$ denotes the transpose. The procedure detailed in \eqref{eq_ista} is applied iteratively until convergence, requiring repeated application of the matrices $\mathbf{A}$, $\mathbf{A}\trans$, and $\mathbf{A}\trans\mathbf{A}$. In practice, these matrices are large and incur a considerable computational and storage cost. 

In a SOTA variant of ISTA, iterations are unrolled into a Deep Neural Network (DNN), where hyperparameters (e.g. step size and thresholds) and even linear mappings are learned from data, thereby improving the effectiveness of each iteration \cite{monga2021algorithm, shlezinger2023model}.
A DNN can be constructed by mapping each ISTA iteration to a block, and stacking multiple such blocks, thereby mimicking the repeated execution of ISTA iterations. Each block, in this context, is defined as a group of layers and activation functions that collectively implement a fixed functionality.
In recent years, the deep unrolled LISTA has been widely studied \cite{gregor2010learning, zhang2018ista, liu2019alista,fu2021structured, wang2024efficient} and tested in various applications \cite{huijben2020learning,xu2024deep,wang2023deep,mulleti2023learning, wang2024jointly}. There have been several types of specialized LISTA architectures that address the challenges of the aforementioned large operators by incorporating prior knowledge and structure into DNNs, achieving more compact models and accelerated training \cite{deng2020model,wang2023data}. Some of these include: Traditional MLP-LISTA \cite{gregor2010learning} uses Multi-Layer Perceptron (MLP) structure that has two fully-connected layers without biases per block, one for $\left(\mathbf{I}-\frac{1}{L}\mathbf{A}\trans\mathbf{A}\right)$ and the other for $\frac{1}{L}\mathbf{A}\trans$. The block-wise parameters count $\mathcal{O}(N_\mathrm{s}^2+N_\mathrm{s}N_\mathrm{d}+1)$, which is either independently trainable or shared across all the blocks to save memory. 
ALISTA \cite{liu2019alista} focuses on deriving an optimal model matrix $\mathbf{A}$ in advance, then only keeping shrinkage-thresholding parameter $\lambda/L$ of each block trainable.
It significantly reduces the number of trainable parameters but at the cost of reduced flexibility and potentially limited recovery performance.
LISTA-Toeplitz \cite{fu2021structured} assumes that Gram matrix $\mathbf{A}\trans\mathbf{A}$ is Toeplitz based on a signal prior, which is efficiently implemented by the convolutional layers. 
In CS case with $N_\mathrm{d}\ll N_\mathrm{s}$, matrix $\mathbf{A}\trans\mathbf{A}\in \real^{N_\mathrm{s}\times N_\mathrm{s}}$ requires more storage than $\mathbf{A}\in \real^{N_\mathrm{d}\times N_\mathrm{s}}$ in spite of its structure. Furthermore, in time delay measurements, it is $\mathbf{A}$ rather than $\mathbf{A}\trans\mathbf{A}$ which exhibits convolutional structure.
BC-LISTA \cite{wang2024efficient} was proposed to construct an efficient fully convolutional architecture by exploring the block-Toeplitz structure of the model matrix $\mathbf{A}$. This structure allows the matrix-based forward model to be analytically reformulated as a series of convolutions.

In addition to the aforementioned structure, DNN-based models can be compressed via methods such as quantization \cite{courbariaux2016binarized}, network factorization \cite{han2015learning}, and knowledge distillation \cite{gou2021knowledge}, which shrink networks by removing weights, lowering numeric precision, or transferring soft labels. However, these methods do not explicitly leverage the linear-operator structure of the forward model. In time delay-based measurements, the forward model exhibits approximately low-rank and convolutional structure that supports structured compression rather than heuristic parameter pruning. 
We therefore pursue structure-aware compression to develop a fast and lightweight forward model for time delay-based measurements that enables efficient SSR algorithms.

Our main contributions are summarized as follows.
First, we implement a compressed forward model by applying OMP \cite{tropp2004greed} and SVD to select a reduced set of Basis Filters (BFs) from the analytic convolutional kernels, and construct an associated mixing coefficients matrix.
Second, we use low-rank CNN decomposition \cite{denton2014exploiting,jaderberg2014speeding, li2019learning} to implement a large convolutional layer as smaller subsequent convolutional layers, where fewer BFs lead to smaller networks.
Third, we use LISTA as a case study to create an architecture that is lightweight, memory-efficient, high-performing, and easy to train: we dub it Compressed BC-LISTA (C-BC-LISTA). By taking multichannel ultrasound imaging as a case study, a comprehensive evaluation supports that the proposed C-BC-LISTA outperforms SOTA baselines in terms of reconstruction accuracy, training efficiency, parameter count, and storage size.
Finally, an ablation analysis indicates that an increased number of blocks and analytic initialization via OMP accelerate training and improve convergence behavior.
\vspace{-0.5em}
\section{Methodology}
\vspace{-0.5em}
\subsection{Slice-Wise Convolutional Forward Model}
\begin{figure}[ht!]
    \centering
    \includegraphics[width=\columnwidth]{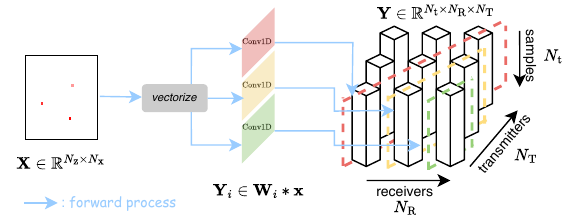}
    \caption{Illustration of the convolutional forward model. }
    \label{fig_conv_forward_model}
    \vspace{-0.2cm}
\end{figure}

We consider a commonly-used time delay-based measurement model that utilizes an $N_\mathrm{c}$-element Uniform Linear Array (ULA), found in applications such as multistatic radar, sonar, medical imaging, and ultrasound nondestructive testing. More concretely, we present Full Matrix Capture (FMC) measurement modality \cite{holmes2005post} in multichannel ultrasound imaging as a representative example of large-scale SSR tasks. As illustrated in Fig. \ref{fig_conv_forward_model}, a specimen with flaws/cracks is described by a sparse reflectivity map $\mathbf{X}\in \real^{N_\mathrm{z}\times N_\mathrm{x}}$. An ULA sequentially transmits pulses, and all elements receive echoes, resulting in a three-dimensional data cube $\mathbf{Y}\in \real^{N_\mathrm{t}\times N_\mathrm{R}\times N_\mathrm{T}}$ with axes [\textit{samples, receivers, transmitters}]. The target localization task is to recover both these scatters' spatial locations and corresponding reflectivity coefficients from the recorded multichannel data. Usually, the number of receivers/transmitters is equivalent to the number of sensor elements $N_\mathrm{c}$.

In this kind of time delay-based measurements, the model matrix $\mathbf{A}$ has a block-Toeplitz structure that can be analytically reformulated as a series of convolutions \cite{wang2024efficient}.
The vectorized reflectivity map $\mathbf{x}\in \real^{N_\mathrm{s}}$ is passed through $N_\mathrm{c}$ parallel convolutional layers, with each layer generating one diagonal slice of the data $\mathbf{Y}$. Owing to the reciprocity principle between transmitters and receivers, two slices positioned symmetrically about the center slice exhibit identical characteristics.
Each slice-wise convolutional forward model is a one-dimensional convolutional layer ($\mathrm{Conv1D}$) with input channel $C_\mathrm{in}=1$, output channel $C_\mathrm{out}=N_\mathrm{t}$, kernel size $K$, stride $S$, and padding $P$. 
A detailed derivation is provided in \cite{wang2024efficient}, and the formulation for a given batch and output channel $i$ is:
\begin{equation}\label{eq_slice_conv}
    y_{i,\tau} = \sum_{k=0}^{K-1}w_{i,k} x_{\tau\cdot S-k+P}.
\end{equation}

\noindent All $C_\mathrm{out}$ filters are stacked in the weight matrix:
\begin{equation}
\vspace{-0.1cm}
    \mathbf{W} =
    \begin{bmatrix}
        \mathbf{w}_1^\mathrm{T} \cdots \mathbf{w}_{C_\mathrm{out}}^\mathrm{T}
    \end{bmatrix}\trans
    \in \mathbb{R}^{C_\mathrm{out} \times K},
\end{equation}
which contains $C_\mathrm{out}\times K$ parameters and the same order of multiplications at inference.
Numerical evidence from previous experiments suggests that typically, some of them are sparse, and many are scaled/shifted versions of one another.
Analytically, for $n$ scatterers, the forward model's time delays form a matrix with rank bounded by $\min(N_\mathrm{c},n)$. Furthermore, the echo-producing pixels for a fixed transceiver pair trace elliptical paths with a sparsity ratio bounded by $2N/N^2$ (assuming low eccentricity and a $N\times N$ grid).
These observations indicate that the weight matrix $\mathbf{W}$ is not full rank, and the repeated use of similar shifts further reinforces this intuition. Consequently, there is a strong possibility that a small subset of basis filters can effectively approximate all kernels through linear combinations.

\vspace{-0.75em}
\subsection{Low-Rank Approximation}
\vspace{-0.5em}
We seek a factorization:
\begin{equation}
    \mathbf{W\approx CB}, \quad \mathbf{B}\in \mathbb{R}^{M\times K}, \, \mathbf{C}\in\mathbb{R}^{C_\mathrm{out}\times M},
\end{equation}
where $\mathbf{B}$ stores $M$ BFs $(M\ll C_\mathrm{out})$ and $\mathbf{C}$ mixes them linearly to reconstruct the original $C_\mathrm{out}$ filters. The goal becomes to find the two optimal matrices that minimize the Frobenius reconstruction error:
\begin{equation} \label{eq_factorization}
    \min_{\mathbf{B},\mathbf{C}} \Vert \mathbf{W-CB} \Vert_\mathrm{F}^2.
\end{equation}
The matrices $\mathbf{C}$ and $\mathbf{B}$ can be obtained via the SVD. However, SVD generally does not preserve the sparsity of the filters. We instead add the constraint that rows of $\mathbf{B}$ must be a subset of the rows of $\mathbf{W}$, turning \eqref{eq_factorization} into a filter interpolation problem. OMP \cite{tropp2004greed} is then used to sequentially select rows from $\mathbf{W}$ to construct the basis $\mathbf{B}$. 
Let $\mathbb{S}$ be the set of selected rows and $\mathbf{R}=\mathbf{W-CB}$ the current residual matrix. For every remaining row $j\notin\mathbb{S}$, we compute its correlation energy with the residual $E_j = \sum_{i=1}^{C_\mathrm{out}} \left\langle \mathbf{R}_i, \mathbf{W}_j \right\rangle^2$,
where $\mathbf{R}_i$ is the $i$-th row of the residual matrix. The row $j^\star = \arg\max_{j\notin\mathbb{S}} E_j$ with the largest energy is selected and added to $\mathbb{S}.$

We repeat this $K$ times and stack the chosen rows $\mathbf{B}=\mathbf{W}_{\mathbb{S}} \in \mathbb{R}^{m\times K}$.
The least-squares optimal mixing coefficients matrix $\mathbf{C}$ that reconstructs $\mathbf{W}$ with this basis is obtained in closed form $\mathbf{C}=\mathbf{W}\mathbf{B}^\mathrm{T}(\mathbf{BB}^\mathrm{T})^{-1}$.
We then compute the new residual matrix $\mathbf{R}=\mathbf{W-CB}$.

We repeat the above steps until the residual is small enough or the number of selected rows reaches a predefined limit $M$. The parameters amount has been reduced from $C_\mathrm{out}\times K$ to $M\times (K + C_\mathrm{out})$, which is a significant compression ratio when $M\ll C_\mathrm{out}$.

\vspace{-0.5em}
\subsection{Low-Rank CNN Decomposition}
\vspace{-0.5em}
\begin{figure*}[ht!]
    \centering
    \subcaptionbox{
        Two-layer slice-wise convolutional forward model.
        \label{fig_slice_conv_decomp}
    }[0.40\textwidth]{%
        \includegraphics[width=0.9\linewidth]{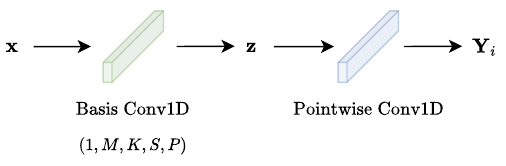}
    }
    \hfill
    \subcaptionbox{
        A single block architecture of C-BC-LISTA.
        \label{fig_compressed_bclista}
    }[0.56\textwidth]{%
        \includegraphics[width=\linewidth]{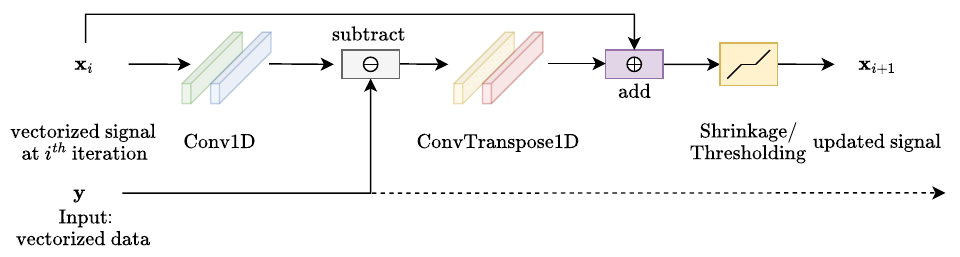}
    }
    \caption{Low-rank CNN decomposition and architecture of C-BC-LISTA.}
    \label{fig:two_subfigs}
    \vspace{-1.5em}
\end{figure*}
Given the basis and mixing matrices, we now show how to realize this factorization as a compact, trainable two-layer convolutional module that is a drop-in replacement for the original layer. For this purpose, we consider low-rank convolutional decomposition for CNN acceleration \cite{denton2014exploiting,jaderberg2014speeding, li2019learning}, except that our BFs are physically interpretable kernels selected from the analytic model rather than learned abstract features. 

Neglecting the batch index, we insert the low-rank expression $w_i=\sum_{j=1}^{M}c_{ij}b_j$ into (\ref{eq_slice_conv}), leading to:
\begin{equation}
    y_{i,\tau} = \sum_{j=1}^{M} c_{ij} \left( \sum_{k=0}^{K-1}b_{j,k}x_{\tau\cdot S-k+P} \right)=\sum_{j=1}^{M}c_{ij} z_{j,\tau},
\end{equation}
where we define basis responses $z_{j,\tau} = \sum_{k=0}^{K-1}b_{j,k}x_{\tau\cdot S-k+P}$. We then have the following two observations. First, computing all $z_j$ for $j=1,\dots,M$ is exactly a 1-D convolution with kernel $\mathbf{B}$, stride $S$, and padding $P$. Second, forming $y_{i, \tau}=\sum_{j}c_{ij}z_{j,\tau}$ mixes channels independently at each time index $\tau$, which is a $1\times 1$ convolution (i.e. a standard matrix multiplication) whose weights are $\mathbf{C}$. 
The slice-wise convolutional layers can then be decomposed into two-layer convolutional modules, as shown in Fig. \ref{fig:two_subfigs}(\subref{fig_slice_conv_decomp}). The first layer is a 1-D convolution with the BFs $\mathbf{B}$, and the second layer is a $1\times 1$ convolution with the mixing coefficients $\mathbf{C}$.

\vspace{-1.0em}
\subsection{Compressed BC-LISTA}
\vspace{-0.5em}
The full forward operator comprises $N_\mathrm{c}$ unique diagonal slices, as shown in Fig. \ref{fig_conv_forward_model}. We analytically select a set of BFs and implement the two-layer decomposition to each slice independently. The resulting compressed forward model approximates the full analytic convolutional forward model while reducing memory. In inverse solvers such as BC-LISTA, the adjoint of the forward model is required. The backward operation simply reverses the two steps and replaces each convolution by its corresponding transposed convolution with the same weights. Because we can copy the weights directly, the adjoint is exact up to numerical precision, so gradients and backpropagation remain faithful to the original model.

We then build the C-BC-LISTA architecture as illustrated in Fig. \ref{fig:two_subfigs}\subref{fig_compressed_bclista}. There are several noteworthy points about customizing the network. First, the illustrated part is just a single block that is algorithmically equivalent to a single ISTA step. The full C-BC-LISTA network is built by stacking multiple such blocks. Second, each block has three sets of parameters: the kernels in the forward model, the kernels in the transposed model, and the shrinkage/thresholding parameter. Each set can be chosen to be trainable or fixed. 
Third, it is possible to share parameters across all the blocks or choose them independently per block. Usually, we prefer to share the kernels but train a unique shrinkage/thresholding parameter for each block.

\vspace{-0.5em}
\section{Evaluation}
\vspace{-0.5em}
\subsection{Training Strategy}
\vspace{-0.5em}
\textbf{Simulation Scenario} $-$ As illustrated in Fig. \ref{fig_conv_forward_model}, the target sparse signal $\mathbf{x}$ is a vectorized reflectivity map containing a random number of scatterers with randomly assigned locations and amplitudes (reflectivity coefficients). 
Although the BC-LISTA and the proposed C-BC-LISTA can be used in large-scale scenarios, such as $128$-channel FMC imaging, the other SOTA algorithms \cite{gregor2010learning, liu2019alista, fu2021structured} are not applicable due to their high memory consumption. To provide a fair comparison, we restrict the number of channels to $N_\mathrm{c}=32$ because of the computational resource limit \footnote{All experiments are conducted on an NVIDIA A100 GPU node.}.

\noindent\textbf{Training Settings} $-$ The training data is synthetic and generated by defining different $(\mathbf{x}, \mathbf{y})$ pairs. 
The data generation strategy is empirically derived from real-world measurements to ensure the model's applicability to real data despite being trained on simulations. To foster robustness against varying sparsity levels, we assume the number of scatterers per map is drawn from a discrete uniform distribution over the integers $5$ to $10$ (i.e., $n\sim \mathcal{U}\{5,\dots,10\}$), every scatterer's reflectivity coefficient follows a Gaussian distribution $a\sim \mathcal{N}(\mu=1250, \sigma^2=250)$. 
This regime allows the model to generalize well under varying SNRs and model mismatches. 
The training data pairs are generated on-the-fly, and the cost function is the Mean Square Error (MSE) between the ground truth $\mathbf{x}$ and the reconstruction $\hat{\mathbf{x}}$.
The training process stops when either an early-stopping criterion is met or the maximal number of epochs ($400$, each consisting of $100$ iterations) is reached. Early stopping is triggered when the validation loss, evaluated on a fixed validation dataset, remains nearly constant ($\Delta\mathcal{L}_\text{val}=(\mathcal{L}_\text{val}^k-\mathcal{L}_\text{val}^{k-1})/\mathcal{L}_\text{val}^{k-1}<\gamma=10^{-6}$) over $p=5$ consecutive epochs, indicating model convergence. 

\noindent\textbf{Training Log} $-$ Following the same training strategy, we trained MLP-LISTA, ALISTA, BC-LISTA, and C-BC-LISTA models with $10$ blocks under both noiseless conditions and noisy environments, with Additive White Gaussian Noise (AWGN) levels of \SI{5}{\decibel}, \SI{10}{\decibel}, \SI{15}{\decibel}, and \SI{20}{\decibel}. All the models can be analytically initialized given these matrices $\mathbf{A}$, $\mathbf{W}$, $\mathbf{C}$, and $\mathbf{B}$.

\begin{figure*}[ht!]
    \centering
    \subcaptionbox{
        Comparison between different number of BFs of 10-layer C-BC-LISTA when SNR = \SI{5}{\decibel}. The predefined epochs are 400, all three settings trigger the early stopping criterion.
        \label{fig_cbclista_val_compare}
    }[0.32\textwidth]{%
        \includegraphics[width=\linewidth]{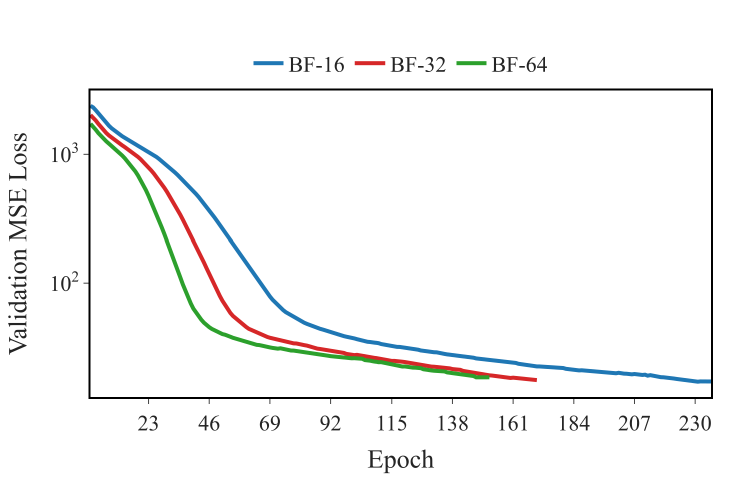}
    }
    \hfill
    \subcaptionbox{
        Comparison across different $10$-layer LISTA models when SNR = \SI{20}{\decibel}. The C-BC-LISTA converges faster than the other models, and achieves the lowest loss.
        \label{fig_val_sota_compare}
    }[0.32\textwidth]{%
        \includegraphics[width=\linewidth]{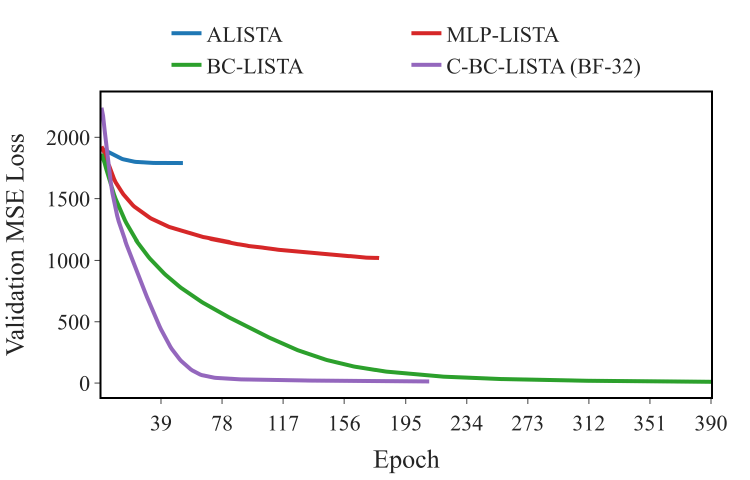}
    }
    \hfill
    \subcaptionbox{
        Ablation comparison at SNR = \SI{20}{\decibel}. Baseline configuration denotes the $10$-layer model initialized by OMP-derived matrices, with both forward and transposed models trainable. 
        \label{fig_ablation_analysis}
    }[0.32\textwidth]{%
        \includegraphics[width=\linewidth]{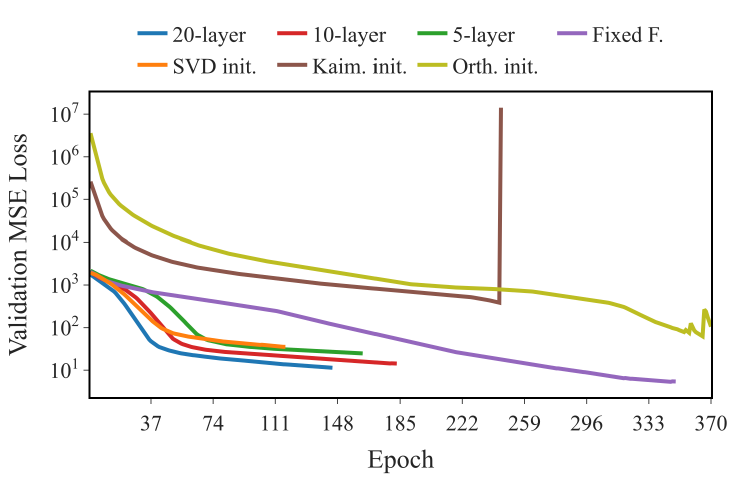}
    }
    \caption{Training validation loss comparison in different settings.}
    \label{fig:three_subfigs}
    \vspace{-0.6em}
\end{figure*}

\noindent The number of original filters of a slice-wise model is $C_\mathrm{out}=N_\mathrm{t}=400$, we have conducted the training of C-BC-LISTA given $M=16,\,32, \,64$ in all the environments. Fig. \ref{fig:three_subfigs}(\subref{fig_cbclista_val_compare}) shows the validation loss comparison between different $M$-values in a noisy environment at \SI{5}{\decibel} SNR level. The three settings finally reach a similar convergence level, but more BFs apparently boosts a faster loss decrease.

\vspace{-1.0em}
\subsection{Quantitative Comparisons}
\vspace{-0.5em}
\begin{table*}[ht!]
    \centering
    \renewcommand{\arraystretch}{1.3}
    \begin{tabular}{lcccccccccc}
        \toprule
        \textbf{Model} & \textbf{Parameters} & \textbf{Storage}  
        & \multicolumn{2}{c}{\textbf{Noiseless}} 
        & \multicolumn{2}{c}{\textbf{SNR = }\SI{20}{\decibel}} 
        & \multicolumn{2}{c}{\textbf{SNR = }\SI{5}{\decibel}} \\
        \cmidrule(lr){4-5} \cmidrule(lr){6-7} \cmidrule(lr){8-9} 
        & &  & PAE(\SI{}{\percent}) & SE & PAE(\SI{}{\percent}) & SE & PAE(\SI{}{\percent}) & SE \\
        \midrule
        MLP-LISTA & \SI{1738}{\million} & \SI{6.96}{\gigabyte} & $\SI{2.48}{\percent}$ & $40.09$ & \SI{2.50}{\percent} & $40.81$ & \SI{2.61}{\percent} & $41.41$ \\
        ALISTA & $10$ & \SI{6.96}{\gigabyte} & \SI{3.36}{\percent} & $19.71$ & \SI{3.39}{\percent} & $20.15$  & \SI{3.36}{\percent} & $19.80$ \\
        BC-LISTA & \SI{81.5}{\million} & \SI{653}{\megabyte} & \textbf{\SI{0.29}{\percent}}  & $0.38$ & \textbf{\SI{0.27}{\percent}} & $0.30$ & \SI{0.35}{\percent} & $0.49$ \\
        C-BC-LISTA (BF-64) & \SI{27.7}{\million} & \SI{106}{\megabyte} & \SI{0.33}{\percent}  & $0.36$ & \SI{0.33}{\percent} & $0.34$ & \SI{0.36}{\percent} & $0.49$ \\
        C-BC-LISTA (BF-32) & \SI{13.8}{\million} & \SI{53}{\megabyte} & \SI{0.32}{\percent}  & $0.33$ & \SI{0.32}{\percent} & $0.31$ & \SI{0.35}{\percent} & $0.46$ \\
        C-BC-LISTA (BF-16) & \SI{6.9}{\million} & \SI{27}{\megabyte} & \textbf{\SI{0.29}{\percent}}  & $\mathbf{0.20}$ & \SI{0.31}{\percent} & $\mathbf{0.20}$ & \textbf{\SI{0.34}{\percent}} & $\mathbf{0.31}$\\
        ISTA-200 & Not Applicable & \SI{6.96}{\gigabyte} & \SI{1.92}{\percent}  & $20.09$ & \SI{1.94}{\percent} & $20.11$ & \SI{1.94}{\percent} & $20.00$ \\
        ISTA-500 & Not Applicable & \SI{6.96}{\gigabyte} & \SI{0.92}{\percent}  & $5.86$ & \SI{0.93}{\percent} & $5.93$ & \SI{0.95}{\percent} & $6.00$ \\
        \bottomrule
    \end{tabular}
    \caption{Performance comparison under noiseless and different noisy environments.} 
    \label{tab:noise_compare}
    \vspace{-1.5em}
\end{table*}
In this part, we provide a comprehensive comparison between the proposed C-BC-LISTA with the traditional ISTA \cite{ista}, MLP-LISTA \cite{gregor2010learning}, ALISTA \cite{liu2019alista}, and BC-LISTA \cite{wang2024efficient}. 

\noindent\textbf{Model Size} $-$ Table \ref{tab:noise_compare} presents a quantitative comparison of the number of trainable parameters and the storage requirements for all LISTA models applied to the $32$-channel FMC measurements. Among all models, the C-BC-LISTA reduces the storage by several orders of magnitude compared to MLP-LISTA. It is important to note that the storage values reported here represent only the model size; actual training typically requires several times more memory due to intermediate computations and gradients.

\noindent\textbf{Convergence} $-$ As illustrated in Fig. \ref{fig:three_subfigs}(\subref{fig_val_sota_compare}), the convergence behaviors of different LISTA models differ significantly. MLP-LISTA shows slow convergence and plateaus at a relatively high loss, while ALISTA maintains a consistently high loss without effective reduction. BC-LISTA performs better, achieving lower loss with stable convergence. Notably, C-BC-LISTA demonstrates the fastest convergence rate and attains the lowest validation loss among all models, highlighting both its efficiency and superior reconstruction capability.

\noindent\textbf{Recovery Performance} $-$ The SSR performance is evaluated by the amplitude fidelity and localization accuracy, which can be quantitatively measured by the Percentual Amplitude Error (PAE) and the Support Error (SE) \cite{rahmathullah2017generalized}. 
The PAE is defined in the worst case by assuming the MSE is caused by a single target, and the SE is the number of pixels, where the true support differs from the estimated support.
The two metrics are computed by:
\vspace{-0.1\baselineskip}
\begin{equation}
    \text{PAE} = \frac{\sqrt{\text{MSE}}}{\mu}\times 100\%,\quad \text{SE}  =  \left| \{ \vert\hat{\mathbf{x}}\vert > 0 \;\oplus\; \vert\mathbf{x}\vert > 0 \} \right|,
\end{equation}
where $\mu=1250$ is the reflectivity coefficient mean and $\oplus$ denotes and logical XOR.
We evaluate all LISTA variants and ISTA baselines (200 and 500 iterations) on the same dataset of 640 pairs—reporting PAE and SE averaged over all pairs—with all models trained using \SI{5}{\decibel} AWGN and evaluated under noiseless, \SI{20}{\decibel}, and \SI{5}{\decibel} test conditions.
The Table \ref{tab:noise_compare} shows that C-BC-LISTA delivers the best overall performance across all conditions; compared with MLP-LISTA and ALISTA, all convolutional variants achieve markedly lower errors with far fewer parameters and smaller storage. 
We observe that increasing the number of filters primarily accelerates convergence speed via over-parameterization rather than lowering the error floor, confirming that the signal subspace is sufficiently captured by $16$ (or fewer) filters.
Consequently, the most aggressively compressed variant, C-BC-LISTA (BF-16), frequently achieves the lowest PAE/SE in several settings.
This indicates that the tight rank constraint acts as a beneficial regularizer, effectively removing redundancy and noise while maintaining superior reconstruction accuracy.

\vspace{-1.0em}
\subsection{Ablation Analysis}\label{subsec_ablation}
\vspace{-0.5em}
The ablation study is conducted to assess the training performance of C-BC-LISTA under different settings, where the analysed factors are the number of blocks, initialization methods \cite{glorot2010understanding, he2015delving, hu2020provable}, and whether the forward model is fixed.
Fig. \ref{fig:three_subfigs}(\subref{fig_ablation_analysis}) presents $7$ loss curves of C-BC-LISTA (BF-$32$). It supports that increasing the number of layers from $5$ to $20$ accelerates convergence and improves final accuracy. Analytic initialization of both models leads to stable and fast convergence, while freezing the forward model slows training. 
SVD has slightly higher computation efficiency than OMP when $M$ is small, and SVD-leading initialization converges at higher loss than the baseline.
Random initialization severely degrades performance: Xavier consistently stalls at high loss, whereas Kaiming and orthogonal initializations cause catastrophic divergence at later stages. These results emphasize that analytic initialization is essential for stable and efficient training of C-BC-LISTA.

\vspace{-0.4em}
\section{Conclusion}
\vspace{-0.2em}
In this paper, we propose a compressed forward model based on low-rank approximation and CNN decomposition. Based on this model, we implement a C-BC-LISTA model and apply it to multichannel ultrasound imaging. 
Quantitative comparisons with SOTA methods demonstrate that C-BC-LISTA achieves competitive performance with significantly fewer parameters, reduced memory requirements, and notably faster convergence and lower training cost. 
The ablation analysis confirms that deeper architectures, analytic initialization, and keeping both the forward and backward models trainable are beneficial for training and inference of C-BC-LISTA.
In future work, we will evaluate these models on real-world measurements and investigate their integration into task-based learning frameworks for CS, with comparisons to task-agnostic approaches \cite{wang2025learning}. Finally, we intend to apply the compressed model to facilitate efficient generative model-based adaptive sensing.

\setlength{\bibitemsep}{0.5pt}
\printbibliography[heading=bibnumbered, title={{REFERENCES}}]    

\end{document}